# Axial Hall Effect in Altermagnetic Lieb Lattices

Xilong Xu,[1] Haonan Wang,[1] Li Yang*, [1,2]

[1]Department of Physics, Washington University in St. Louis, St. Louis, Missouri 63130, USA

[2]Institute of Materials Science and Engineering, Washington University in St. Louis, St. Louis, Missouri 63130, USA

**Abstract**

We predict a so-called axial Hall effect, a Berry-curvature-driven anomalous Hall response, in Lieb-lattice altermagnets. By constructing a tight-binding model, we identify the axial direction as a hidden topological degree of freedom. Breaking the double degeneracy of axial symmetry generates substantial Berry curvature and induces a pronounced anomalous Hall conductivity. First-principles calculations further confirm the emergence of this effect in strained altermagnets, particularly in ternary transition-metal dichalcogenides. We take $Mn_2WS_4$ as an example to reveal that the axial Hall effect originates from the interplay between Dresselhaus spin-orbit coupling and the intrinsic piezomagnetic response of Lieb-lattice altermagnets, leading to highly localized and enhanced Berry curvature. Remarkably, the magnitude of the axial Hall effect is significant and remains unchanged when varying the strain, highlighting the topological nature of the axial degree of freedom. Finally, in multilayer systems, the effect manifests as a distinctive thickness-dependent modulation of both anomalous and spin Hall responses. These findings emphasize the critical role of spin-orbit coupling and noncollinear spin textures in altermagnets, an area that has received limited attention, and open new pathways for exploring intrinsic Hall phenomena in topological magnetic systems.

The Hall effect, discovered by Edwin Hall in 1879, and subsequently anomalous, spin, and quantum Hall phenomena hold a central-importance position in condensed matter physics. [1–15] The emergence of two-dimensional (2D) materials has profoundly enriched the Hall-effect landscape by introducing various new degrees of freedom (DOF). [16–20] Notable examples include the recently proposed valley Hall effect and layer Hall effect. [21–24] Both originate from intrinsic contributions governed by the electronic Berry curvature with the help of spin-orbit coupling (SOC). [22–26] The valley Hall effect is characterized by the valley pseudospin DOF, with valleys connected via time-reversal symmetry. It can be detected by breaking the time-reversal symmetry via an external magnetic field or circularly polarized light. [21,27–31] Meanwhile, the layer Hall effect involves the layer pseudospin DOF and is sensitive to inversion or mirror symmetry, which can be broken by applying an electric field to induce a detectable anomalous Hall conductivity (AHC). [22,32–34] These new DOF significantly enhance the functional capabilities of 2D quantum materials because they provide novel pathways to control charge and spins for information encoding, quantum transport manipulation, and device miniaturization. [35–38]

Altermagnetism is a recently identified magnetic phase in addition to the traditional classification of magnetism, ferromagnetic and antiferromagnetic types. [39–49] Unlike ferromagnets, which possess a finite net magnetization, and antiferromagnets, in which opposite magnetic moments generally cancel out to preserve Kramers' degeneracy due to the *Translation+T* or *PT* symmetry, altermagnets exhibit zero net magnetization yet display symmetry-enforced spin polarization patterns in reciprocal space because of the interplay between crystal rotational symmetry and magnetic ordering. [39,40] As a result, altermagnets can support numerous unique properties, such as alternating chiral magnons, multiferroics with alter-spin-ferroelectric locking, and altermagnetic magnon-mediated superconductivity. [50–61] Among these intriguing properties, a fundamental while promising one is the spin-dependent transport phenomenon, even without macroscopic magnetization, such as AHC. However, this AHC arises from extrinsic scattering, and its magnitude is relatively weak in altermagnets. [8,62] Given the importance of Berry-curvature driven Hall effects in fundamental science and applications, there is broad interest in identifying intrinsic Berry curvature, strong anomalous Hall responses, and, more deeply, understanding the role of SOC and noncollinear spin texture, a field which has been overlooked in altermagnets.

In this work, we propose to realize a pronounced anomalous Hall response arising from large, localized Berry curvature in altermagnetic Lieb lattices and introduce the axial Hall effect, characterized by a new pseudospin DOF, the axial direction. Using a tight-binding model incorporating SOC, we explore the conditions for generating substantial intrinsic Berry curvature and AHC in $S_4T$-symmetric altermagnetic Lieb lattices. First-principles calculations demonstrate the axial Hall effect in ternary transition-metal chalcogenides, such as $Mn_2WS_4$, under applied uniaxial strain. The interplay between piezomagnetic response and Dresselhaus SOC generates highly localized Berry curvature and intrinsic AHC. Importantly, the axial Hall response emerges abruptly, and its magnitude remains unchanged by the strength of the applied strain. These characteristics highlight the axial direction as a new topological DOF: similar to the valley Hall or layer Hall effect, the strain breaks the axial symmetry and exposes the hidden DOF. Finally, we extend this concept to few-layer altermagnetic Lieb lattices, predicting layer-dependent anomalous and spin Hall responses. The axial DOF and corresponding Hall effect underscore the significance of SOC in altermagnets and open promising avenues for future Hall-effect-related applications.

Altermagnetic Lieb Lattice: Lieb lattices have garnered extensive research interest due to their distinctive flat band and topological properties. As illustrated in Fig. 1(a), the Lieb lattice can be described by three sites in a square unit cell. Two of the sites (A and B) are on the edge of the unit cell. The third site (O), which has four neighbors, is at the corner. If we put the opposite spin vector on the edge sites (A and B), the $C_4T$ ($S_4T$) joint symmetry will appear, which may form an altermagnetic symmetry. The corresponding tight-binding Hamiltonian is [63,64]

$$H_0 = -\mu \sum_i c_i^* c_i - t_1 \sum_{<i,j>} c_i^* c_j - it_2 \sum_{\ll i,j \gg} v_{ij} c_i^* c_j + m \sum_i l_i c_i^* \sigma_z c_i \qquad (1)$$

where $\mu$, $t_1(t_2)$, $m$ and $\sigma_z$ represent the on-site energy of site $i$, the (next) nearest-neighbor hopping, the magnetic moment, and spin Pauli matrix, respectively. The $l_i$ is the axial index. It is +1 along the x-axis chain (A site) and -1 along the y-axis chain (B site), distinguishing between degenerate yet distinct sublattice atomic sites. The corresponding band structure is plotted in Fig. 1(b), and the characteristic flat bands of Lieb lattices are observed around the Fermi level. Importantly, the electronic bands exhibit alternative spin splitting along the Γ-X-M and Γ-Y-M paths, while they remain degenerate along the high-symmetry Γ-M path, resulting in a typical *d*-wave altermagnetism. (See the detailed spin distribution of the first Brillouin zone in Fig. S2).

Notably, the valence band maximum (VBM) at the M point mixes the spin-up and spin-down components. Thus, it is necessary to further include momentum-dependent SOC to clarify the fine spin structure there. To simplify the analysis, we begin with the Dresselhaus SOC, which excludes long-range electric-field effects. This limits our discussion for non-polarized materials or external electric field. The SOC Hamiltonian near the M point is:

$$H_D = \lambda_D \left(\sigma_x k_x - \sigma_y k_y\right), \qquad (2)$$

where $\lambda_D$ represents the strength of Dresselhaus SOC. Combining Eq. (2) with Eq. (1), we can obtain the noncolinear spin distribution of the first valence band shown in Fig. 1(c). Most regions predominantly keep a pure spin polarization along the z direction, implying that the fundamental *d*-wave altermagnetic order remains intact.

On the other hand, significant spin mixing occurs near the Γ-M path due to SOC. We further perform a perturbative expansion near the M point (VBM), yielding the local band dispersion as illustrated by dash lines in Fig. 1(d), which are essentially Rashba-like band dispersions, (see Fig. S3 for the detailed spin texture). Thus, the corresponding Hamiltonian around the M point is described by:

$$H_M(\boldsymbol{k}) = \frac{\hbar^2 k^2}{2m^*} + L_x \left(\lambda_D(k_x s_x - k_y s_y)\right) + L_z m_z s_z \;. \qquad (3)$$

*L* is Pauli matrix, which has the same formula as the valley pseudospin in valley Hall and the layer pseudospin in layer Hall effects. [24,32,65] Thus, in this altermagnetic system, *L* represents a new pseudospin DOF, which is defined as the axial DOF. It connects the Dresselhaus SOC terms and magnetic moment with axial directions.

Notably, previous works showed that opening a gap of the Dirac point of the Rashba band dispersion can introduce Berry curvature and nontrivial topological properties. [62,66] In this tight-binding model, if we slightly modify two opposing magnetic moments in two axial directions (sublattices A and B), it is equivalent to imposing a small net magnetic moment. As illustrated in Fig. 1(d), the electronic band exhibits a small pseudo-gap at the original band crossing (M point) in this situation. As a result, our tight-binding calculation confirms that a nonzero and significant Berry curvature emerges (Fig. 1(e)). Importantly, SOC is crucial for this mechanism: without the Dresselhaus term, no gap opens, and the Berry curvature is nearly negligible. (See details in Fig. S4)

Figures 1(e) and 1(f) show that the Berry curvature switches the sign when reversing the relative magnetic moments of axial sublattices A and B. This Berry curvature distribution is anisotropic, with its orientation corresponding to the defined axial index (axial DOF). Based on this tight-binding model, the emergence of a nonzero Berry curvature implies that an intrinsic AHC response may occur. Because of its origin from the axial DOF, the corresponding AHC is so called the axial Hall effect.

In summary, the following requirements have to be satisfied to observe the proposed axial Hall effect in materials: *Non-centrosymmetric altermagnetic Lieb lattices and breaking the degeneration of spin sublattices.*

Altermagnetic $Mn_2WS_4$: We propose a family of 2D van der Waals (vdW) ternary transition-metal chalcogenides, $A_2MX_4$ (A = Mn/Fe, M=W/Mo, and X = S/Se/Te) as the candidates to realize the proposed axial Hall effect. Their basic atomic structure is shown in Fig. 2(a) and corresponding structure parameters are summarized in Table S1. A few members of these materials, such as $Cu_2WS_4$, and $Ag_2WS_4$, have been successfully synthesized and exfoliated. [67–74] In addition, Mo-based materials with the same structure, such as $Fe_2MoX_4$ (X = S/Se/Te), have also been investigated. [75–78] In the following, we take $Mn_2WS_4$ as an example, and other materials in this family exhibit the similar axial Hall effect. This family of materials belongs to the space group *P-4m2*, with confirmed stability (see details in Section II of SI). First-principles DFT+U calculations (see details for Section III in SI) show each manganese ion carries about 4-$\mu_B$ magnetic moment with tungsten and sulfur atoms being nonmagnetic. The ground magnetic states exhibit an antiferromagnetic (AFM) tetragonal structure. This results in the formation of a combined spin symmetry group, $\{C_2\|S_4\}$, which ensures a $S_4T$ symmetry and that the bands are spin-split with opposite spins at the high-symmetry X and Y points in reciprocal space, and a distinct *d*-wave altermagnetic structure.

First, without SOC, the band structure is shown in Fig. 2(b). Its main features are like the tight-binding results plotted in Fig. 1(b) and exhibit the altermagnetism: the band dispersions along Γ-X and Γ-Y directions are energetically degenerate but with opposite spins. Notably, the spin splittings are several hundred meV for both conduction and valence bands. Such significant bipolar spin splittings are preferred for experimental detection and spintronic applications.

Figure 2(c) presents the band structure with SOC included. It is similar to the non-SOC result in Fig. 2(b), except for a noticeable transverse band splitting along the Γ-M path and that VBM is more prominent at the M point. Because there is no net electric polarization, the non-centrosymmetric nature of this material will manifest itself in the form of Dresselhaus SOC, as supported by the plotted spin texture of the first valence band in Fig. 2(d). Like the tight-binding model, there is only in-plane spin component but no out-of-plane spin component in the nearby area along the diagonal direction (Γ-M) in the reciprocal space. In other regions, the out-of-plane component still dominates, keeping the basic spin-band characteristics of altermagnetism.

Breaking the degeneracy of spin sublattices: The altermagnetic state with net-zero magnetic moment originates from the presence of sub-symmetry of $M_{xy}$. Thus, breaking these symmetries will lift the degeneracy between two magnetic sublattices and result in net magnetic moments. We find that the uniaxial strain is a straightforward way to achieve this goal. In fact, there have been several works reporting the piezomagnetic effect in altermagnets, [53,79–81] including our recently predicted giant piezomagnetism in these ternary transition-metal chalcogenides. [82] For monolayer $Mn_2WS_4$, the variation of the net magnetic moment per unit cell under applied axial stress ranging from -5% to 5% is shown in Fig. 3(a).

Taking +2% strain as an example, in which the strain-induced net magnetic moment is about 0.007 μ$_B$/unit-cell. The corresponding band structure is plotted in Fig. 3(b). Crucially, as predicted in Fig. 1(d) by the tight-binding model, the imbalance spins between two sublattices (piezomagnetic effect) induce a narrow gap at the M point (See the inset of Fig. 3(b) and details of spin textures in Fig. S9). The corresponding Berry curvature emerges: the inset of the upper panel of Fig. 3(c) shows the Berry curvature of the first valence band for +2% strain along the x direction. Agreeing with the tight-binding model (Fig. 1(f)), the Berry curvature primarily located around M point and indicates a highly anisotropic shape along the y direction (vertical to the strain direction). Conversely, if we apply a +2% stress along the y direction, as shown in the inset of the lower panel of Fig. 3(c), the sign of Berry curvature reverses, and its distribution in reciprocal space is rotated by 90°.

It is worth noting that these Berry curvature distributions are not from the net magnetic moment induced by piezomagnetism, but from the broken degeneracy of the intrinsic opposite Berry curvatures of Dresselhaus SOC. As shown in Fig. 3(c), when we turn on strain, the Berry curvature suddenly appears. Although the peak value of the Berry curvature changes as increasing the strain, the distribution of Berry curvature expands, and its integral remains nearly unchanged.

Axial Hall effect: Figure 4(a) plots the spectra of calculated intrinsic AHC for unstrained and strained monolayer $Mn_2WS_4$. (See Section V of SI for calculation details) Without strain, the intrinsic AHC is zero due to the $S_4T$ symmetry. While with +2% strain along the x direction, significant AHC emerges. Take the example of hole doping. An AHC peak A is located around 60 meV below the Fermi level. The peak value is significant (~ 0.3 $e^2/h$), and the AHC signal can be detected over a wide hole-doping range (150 meV below the Fermi level). These two features are preferred for measurements and applications.

The variation of the peak A value of AHC with the magnitude of strain is plotted in the upper panel of Fig. 4(b) (details in Section V of the Supplementary Information). Notably, the AHC signal appears abruptly upon applying axial strain but remains constant as the strain increases. This supports the intrinsic nature of the axial DOF, as the axial strain merely breaks the axial degeneracy and exposes one of the components. The nearly constant AHC is consistent with the nearly constant integral of Berry curvature results shown in Fig. 3(c).

We have also calculated the width of the AHC peak A associated with the strain. As shown in the lower panel of Fig. 4(b), the full width at half maximum (FWHM) increases by increasing the magnitude of the axial strain. Such a wider FWHM makes it easier for measurements of the predicted axial Hall effect.

Figure 4(c) presents the proposed axial Hall effect schematically. When the y-direction tensile strain is applied, doped holes are driven by the nonzero in-plane Berry curvature to deflect to the left edge, leading to AHC. Particularly, because of the piezomagnetic effects, the VBM is slightly spin-up polarized. Thus, the observation will include both anomalous Hall and spin Hall effects. Conversely, when the y-direction compressive strain is applied, doped holes deflect to the right edge, resulting in opposite charge and spin AHC. This also agrees with Fig. 3(c), in which the switch of the strain from the y direction to the x direction reverses the sign of Berry curvature. Table I summarizes the axial responses of hole doping for both axial and diagonal strains. All these sensitivities of AHC effects for the sign and direction of strain demonstrate the strong coupling between this Hall effect and the crystal axis DOF, justifying the term of the axial Hall effect. (The partial charge density near the valence band maximum under applied strain is presented in Fig. S10 in SI)

Beyond monolayers, we have also investigated bilayer and multilayer configurations. When two layers are stacked, the ground-state interlayer AFM ordering results in two sets of spin-opposite $S_4$ symmetry operations. Upon applying axial strain, a pseudo bandgap still forms near the M point (See Section VI in SI). As illustrated in Fig. 4(d), each individual layer contributes to an axial AHC, but two layers exhibit opposite AHC responses. Overall, they effectively cancel out the net charge Hall signal but result in a pure spin Hall conductivity. Intriguingly, the axial DOF will reverse charge and spin distributions on each layer by switching the type of applied strain (tensile vs. compressive), which enriches the layer-related Hall response. When extended to trilayer, the net axial charge Hall response re-emerges due to the residue layer. Consequently, the system exhibits an alternatively layer-number-dependent axial Hall response: *odd-numbered layers support an axial anomalous Hall effect, while even-numbered layers support an axial spin Hall effect*.

Finally, Table II compares different DOF corresponding to distinct Hall effects: (1) Valley DOF, represented by two valleys, which have opposite intrinsic Berry curvature. The valleys are connected through time-reversal symmetry and can be distinguished via magnetic doping or circularly polarized light, enabling the detection of the valley Hall conductivity. Notably, an interesting valley Hall effect has been recently predicted in altermagnets at the X/Y valleys recently. [75,77,83] (2) Layer DOF, represented by the top and bottom layers, which carry opposite Berry curvatures. These layers are connected through spatial inversion or mirror symmetry and can be distinguished via an applied electric field, allowing the detection of the layer Hall conductivity. (3) Axial DOF, represented by the x-axis and y-axis, which carry opposite Berry curvature and are connected via the $S_4T$ symmetry. This DOF can be exposed by applying axial strain, which breaks the $S_4T$ symmetry, enabling the measurement of the charge and spin axial; Hall conductivity, as illustrated in Figs. 4(c) and 4(d).

**Acknowledge:**

X.X. is supported by the National Science Foundation (NSF) Designing Materials to Revolutionize and Engineer our Future (DMREF) DMR-2118779. L.Y. is supported by NSF DMR- 2124934. The simulation used Anvil at Purdue University through allocation DMR100005 from the Advanced Cyberinfrastructure Coordination Ecosystem: Services & Support (ACCESS) program, which is supported by National Science Foundation grants #2138259, #2138286, #2138307, #2137603, and #2138296.

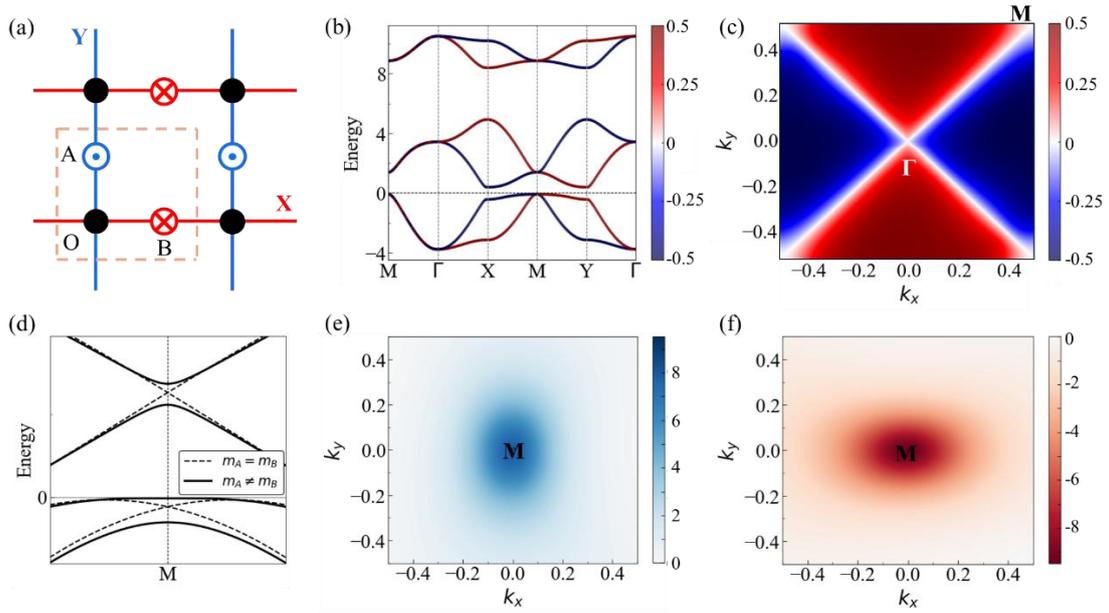

**Fig. 1**. (a) Schematic representation of the altermagnetic Lieb lattice structure, showing its geometrical and magnetic arrangement. (b) Tight-binding band structures for the altermagnetic Lieb lattice without SOC. The color coding in the figure reflects the spin polarization. The Fermi level is set to be zero. (c) Plot of the $S_z$ expectation values for the first valence band across the first Brillouin zone, showcasing the distribution of spin polarization in momentum space with SOC included. (d) Band structures derived from a k·p model around the M point with $m_A = m_B$ and $m_A \neq m_B$. (e) and (f) Berry curvature distributions for the first valence band for $m_A = m_B$ and $m_A \neq m_B$, respectively.

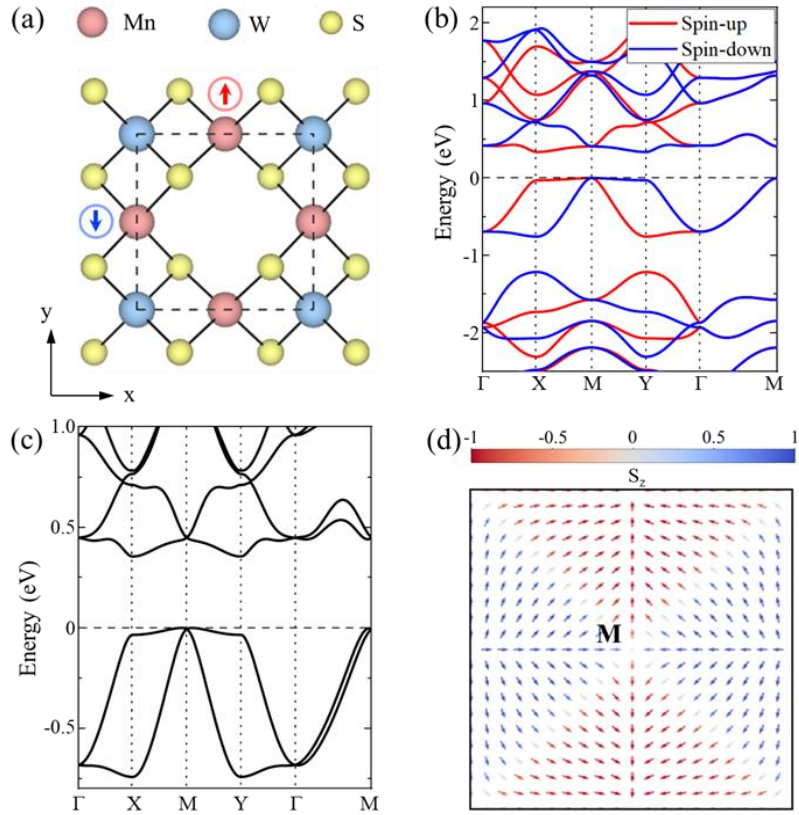

**Fig. 2**. (a) Top view of monolayer $Mn_2WS_4$. (b) Band structures of $Mn_2WS_4$ without SOC. (c) Band structures of $Mn_2WS_4$ with SOC. (d) Spin texture of the first valance band in the first Brillouin zone of $Mn_2WS_4$. The arrow indicates the in-plane spin component ($S_x$, $S_y$), and the color indicates the out-of-plane spin component ($S_z$).

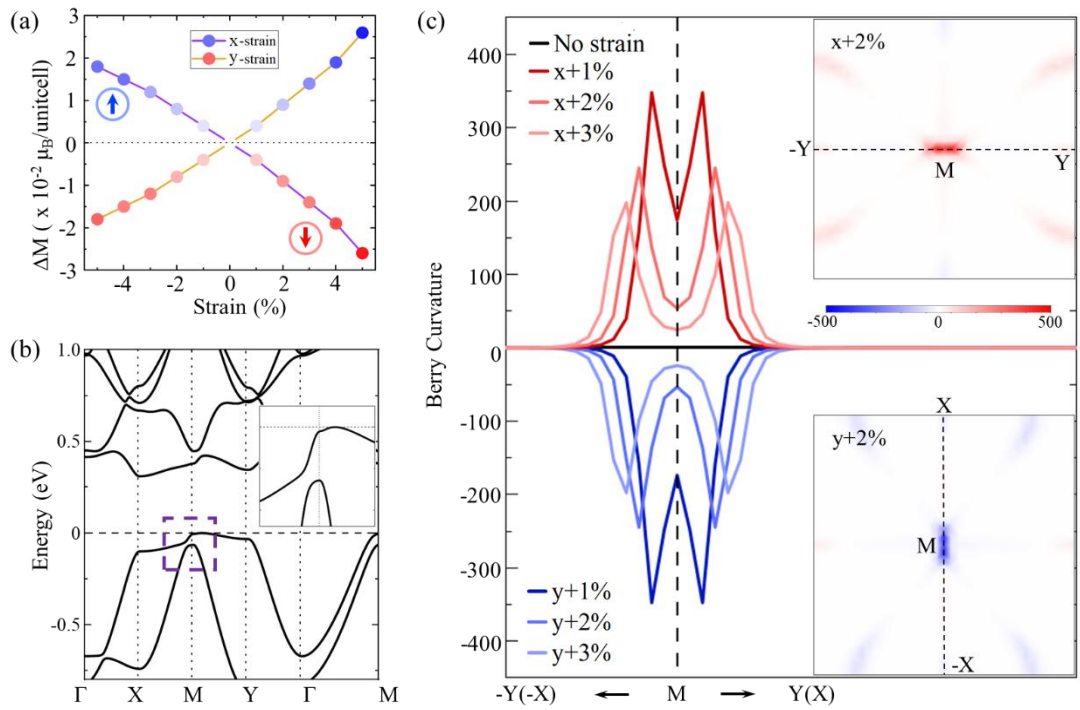

**Fig. 3**. (a) Piezomagnetic effect in monolayer $Mn_2WS_4$. (b) Band structures of $Mn_2WS_4$ with SOC under y+2% strain. (c) Berry curvature distributions of the first valence band of $Mn_2WS_4$ vs strain.

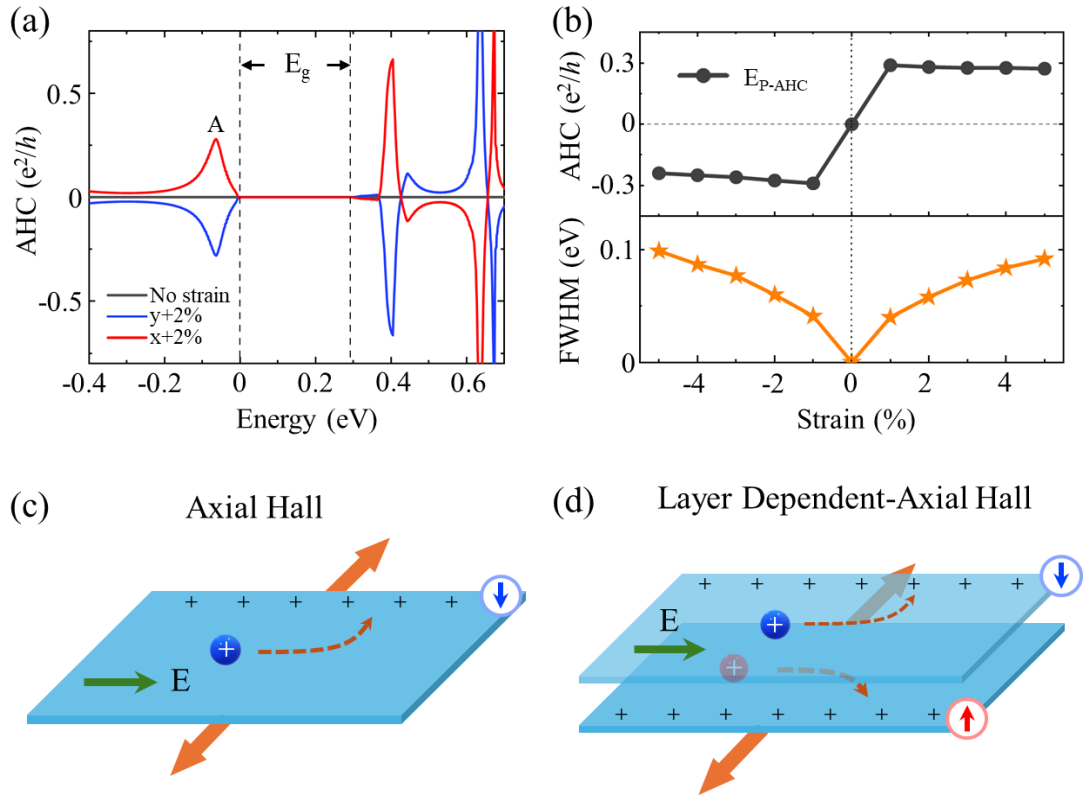

**Fig. 4**. (a) Anomalous Hall conductivity of strained $Mn_2WS_4$ around the Fermi level. (b) Evolution of AHC and AHC-FWHM of monolayer $Mn_2WS_4$ with strain. (c) Schematic of the axial Hall effect of $Mn_2WS_4$ with strain (in orange color). (d) The schematic of layer dependent axial Hall effect of bilayer $Mn_2WS_4$ under strain. The circled arrow stands for the spin accumulation.

**Table 1**. Deflection direction of holes with uniaxial strains in monolayer $Mn_2WS_4$. (The deflection of electrons is opposite to that of holes)

| Sign \ Direction | x | y | diagonal |
|---|---|---|---|
| Tensile | Right | Left | 0 |
| Compressive | Left | Right | 0 |

**Table 2**. Berry-curvature driven Hall effects.

| DOF | Symmetry | Characters | Experimental Observation |
|---|---|---|---|
| Valley | T | 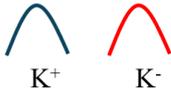 $K^+$  $K^-$ | 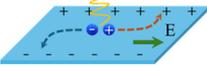 Circular-Polarized |
| Layer | $PT/M_zT$ | 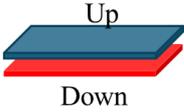 Up / Down | 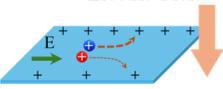 Electric Field |
| Axial | $S_4T$ | 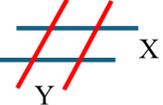 X / Y | 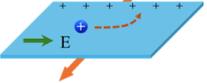 Strain |